\documentclass{article}
\usepackage{graphicx} 
\usepackage{subcaption}
\usepackage{float}
\usepackage{tikz}
\usepackage{amsmath}
\usepackage{booktabs}
\usepackage{pifont}
\usepackage{filecontents}
\usepackage{adjustbox}

\usepackage{xspace}
\usepackage{xcolor}
\usepackage{amssymb}
\usepackage[inline]{enumitem}
\usepackage[table]{xcolor}
\usepackage{hyperref}
\usepackage{xurl}
 \usepackage{tcolorbox}

\usetikzlibrary{arrows.meta,positioning,patterns}

\newtheorem{theorem}{Takeaway}

\title{From Alignment to Access Control: A Framework for GenAI Policy Enforcement}
\author{Nathalie Baracaldo \\ IBM Research }
\date{September 2026}

\begin{document}

\maketitle

\begin{abstract}

    Generative AI (GenAI) applications have flourished enabling users to chat with large language models, and to create agents to act on their behalf for a variety of tasks.
    The pace of development of capabilities in this field is incredibly fast
    with security and safety taking a back seat. Unfortunately, the slower pace at which security and safety mechanisms have evolved has led to real incidents. 
    Policy enables the definition of desirable behavior of applications, and for that reason, it is a cornerstone of making systems secure and compliant.
    \textit{Policy} however means different things to different practitioners creating confusion and siloed solutions that are not adequate for compliance.
    This paper takes a tour of the good, the bad and the ugly when it comes to
    policy enforcement in GenAI applications.
    We propose a methodology to systematically analyze and dissect existing approaches to define and enforce policy found in the wild. 
    Based on this principled analysis, we provide recommendations and call for action for the community to address.

    \vspace{0.5em}
    \noindent
    This paper is a companion extension of USENIX Security 2026 Enigma talk titled \textit{``From Alignment to Access Control: A Unified View of GenAI Policy Enforcement"} by the author \textit{Nathalie Baracaldo}.

\end{abstract}

\section{Introduction}

With the fast evolution in capabilities of large language models (LLMs)
a great amount of Generative AI (GenAI) applications have emerged in a variety of sectors including finance \cite{anthropic:fiance:agent2026}, vibe coding \cite{vibecoding}, information technology (IT) automation tasks
 \cite{itbench-trajectories-2025},
companion chat bots \cite{characterAI}, among many others.
Their power has enabled users to use agents to run workflows and tasks on behalf of users and organizations, alike. Without a doubt, the power of LLMs have opened a new era of opportunities \cite{wang2024surveyLLMAgents,xi2025riseLLMAgentSurvey,pan2025measuringAgentsMarquita}.

\begin{figure}
    \centering
    \includegraphics[width=0.5\linewidth]{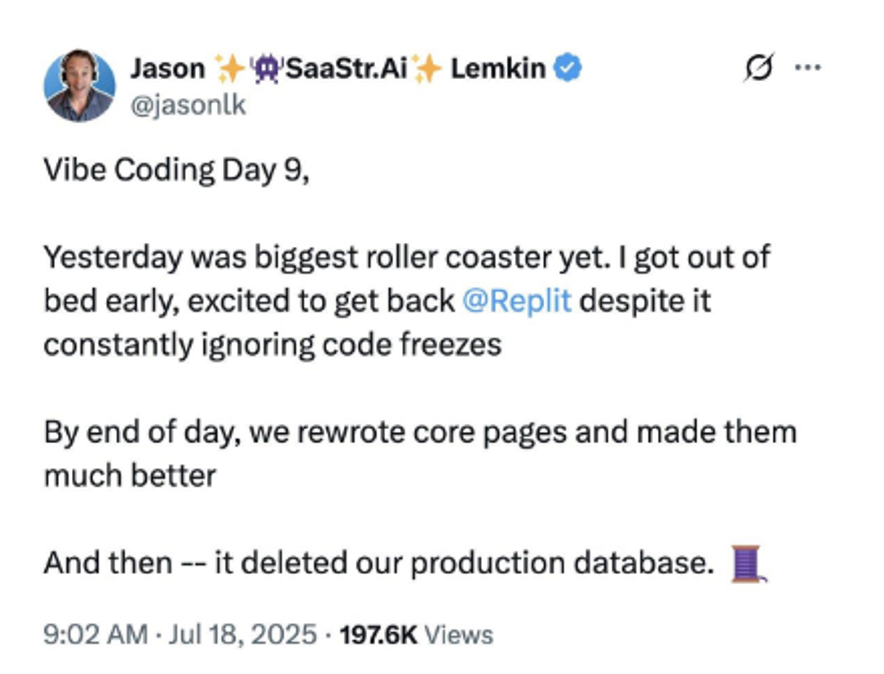}
    \caption{Screenshot depicting vibe coding incident that resulted in a production database deletion \cite{deletion-insident}. }
    \label{fig:incidents}
\end{figure}

At the same time, new incidents have impacted organizations and people in negative ways \cite{he2025securityThreatAgentSurvey,felonyBench}. We highlight two incidents that are different in nature. The first incident occurred while vibe coding, where a GenAI agent deleted a production dataset despite receiving clear instructions not to make any changes \cite{deletion-insident} (Fig. \ref{fig:incidents}).
The second type of incidents took place
when the interaction with LLMs led users to delusional thinking that caused them to endanger their lives or the lives of others \cite{incident:sycophancy:WallStreet,moore2026characterizingdelusionalspiralshumanllm}. 

What do these incidents have in common?
At first glance, not much. One is related to cybersecurity while the other one is related to psychosis generated by interacting with an LLM.
However, after further inspection, the incidents do have one thing in common: they could have been prevented (or at least mitigated) if the right policies were defined and enforced. In the first incident,  having adequate access control enforcement and 
in the second one, by steering the model away from sycophantic spirals. 
But, how should these policies be defined and enforced?

\begin{figure}
    \centering
    \includegraphics[width=0.9\linewidth]{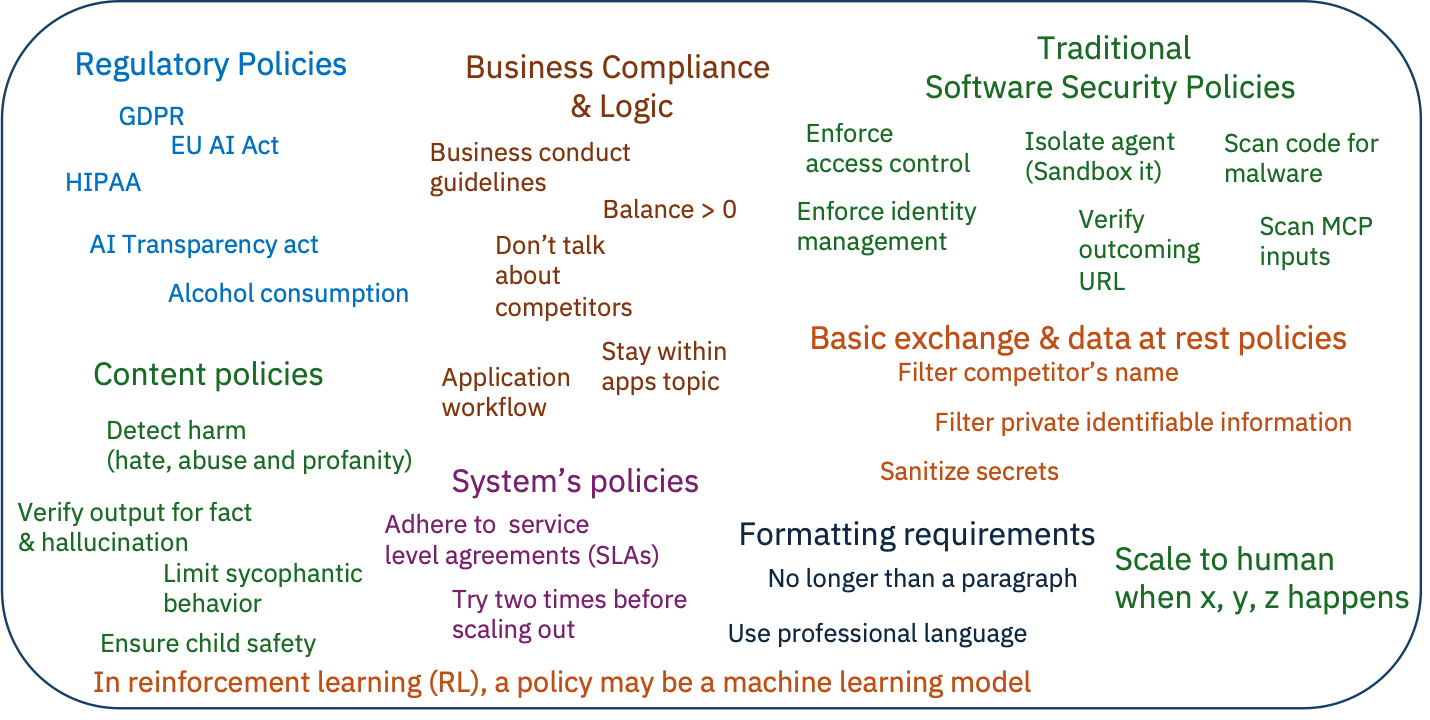}
    \caption{The word \textit{policy} has a diverse set of meanings. A single GenAI application needs to be governed by multiple such policies. Discovering, specifying and enforcing such a diverse set of policies is an open challenge.}
    \label{fig:policy-clusters}
\end{figure}

Before answering that question, let's define what we mean by GenAI policy. The word ``policy" itself carries radically different meanings across communities as illustrated by Fig.  \ref{fig:policy-clusters}. Security practitioners think in terms of mandatory access controls and formal rule systems. Application developers treat policy as agent flow constraints and output validation. 
Policies can also contain service level agreements (SLAs) or other types of application level requirements, for example the minimum balance an account can have, or in other cases, user enrollment requirements. 
AI researchers grapple with softer, high-level behavioral norms: for example, maintaining a non-sycophantic tone, where even the definition of compliance is fuzzy. In the reinforcement learning jargon a machine learning model that has learned certain rules is referred to as \textit{policy}.

As the list shows, even agreeing on what we mean by policy can be challenging.
To understand this tension, let us analyze how different communities define and enforce policies in different ways.
For the cybersecurity community, defining and enforcing policy is a cornerstone of the field with formal languages to specify policy and enforcement mechanisms with verifiable properties.
Unlike conventional cybersecurity systems with well-defined enforcement boundaries, GenAI applications operate in a highly dynamic, probabilistic environment, making policy enforcement a non-trivial, interdisciplinary challenge.
The non-deterministic nature of GenAI makes it incredibly useful, but at the same time introduces a fundamentally new and evolving threat landscape where 
existing mechanisms fall short. 
Not all the desired policies can be defined and enforced in a deterministic fashion.

What's more, with the evolution of the LLM capability, the possibility of specifying a policy in plain natural language has made it possible to define fuzzy policies that were un-enforceable in the past. This has led to a variety of novel approaches for policy enforcement that, as we will see, may generate a false sense of security and compliance.

In addition, different communities often operate in silos, producing enforcement approaches that are fragile, ad hoc, and critically, designed to handle only one policy at a time. Real enterprise GenAI applications, however, demand simultaneous compliance with multiple, overlapping policies across all these layers -- a requirement today’s tools are not built for.
Until the community addresses this fragmentation and acknowledges what solutions are problematic, true compliance in enterprise AI systems will remain an illusion.

In this paper, we face all these dichotomies to provide a much needed 360 view of the field.
We propose a methodology to analyze state of the art approaches for policy specification and enforcement, and understand what defense in depth means for different deployments.
Our analysis brings to light the necessity of having solutions that encompass a diverse set of policies in a single control plane to ensure governance.

Our \textbf{contributions} are as follows: 

\begin{itemize}
  
    \item \textit{Tour of the Wild:}
    We survey what different communities of practitioners refer to as ``policy", what type of interactions they aim to govern and the mechanisms that have been proposed to enforce such policies.

    \item \textit{A Framework to Dissect GenAI Policies:}
    Given the diverse set of approaches found in the wild, it is a cumbersome task to discern their appropriateness and establish commonalities. To close this gap, we propose a methodology to determine the appropriateness of competing policy enforcement mechanisms.
    The methodology dissects solutions based on three factors: policy definition, policy enforcement guarantees and stack enforcement.   
    Our approach aims to help guide the community in the selection of the adequate policy enforcement mechanisms for a diverse set of policies applicable to GenAI applications.

    \item \textit{Open Challenges and Call to Action:}
    Through this tour of the wild, we conclude that there is no single enforcement mechanism for GenAI policy today: policy is fragmented across model alignment, runtime guardrails, agent flow constraints, and access controls, with no unified view across layers. Some clearly flawed approaches are used out there.
    We conclude the paper by providing recommendations to avoid common pitfalls and discussing open challenges. 

\end{itemize}

The rest of the paper takes a top-down approach starting by highlighting the trends in GenAI policies in the wild (Sec. \ref{sec:policies-in-the-wild}) and presenting the proposed methodology to dissect them (Sec. \ref{sec:approach-to-dissect-policies}). As we unfold different findings, we highlight takeaways of our analysis.
We provide more information about existing approaches in Sec. \ref{sec:sota}.
Finally, in Sec. \ref{sec:conclusions}, we conclude the paper by summarizing the recommendations and open challenges.
We hope this work will enable the community to generate safer and compliant solutions.

\section{Policies in the Wild} \label{sec:policies-in-the-wild}
Policy is a somewhat fuzzy word that may mean different things to different audiences.
This overloaded word results in difficult communication among communities and how they perceive and propose solutions for GenAI compliance.
The diversity of policies is also due to the very diverse set of applications and users of LLM-based systems. We start by defining basic concepts and providing aggregated information of our survey findings. Detailed information about each approach is available in Section \ref{sec:sota}.

\begin{figure}
    \centering
    \includegraphics[width=0.75\linewidth]{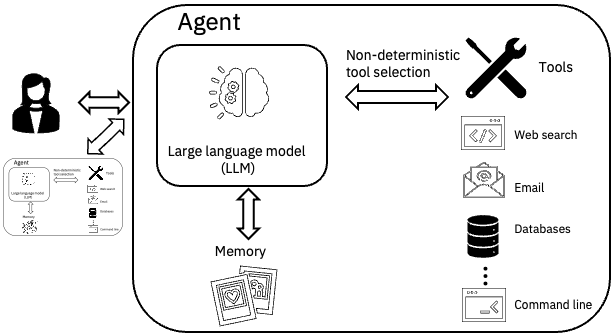}
    \caption{A typical GenAI agent}
    \label{fig:agent}
\end{figure}

\textbf{What is a GenAI Application?}
We use the term GenAI applications to refer to
\textit{agents} that use LLMs to complete tasks or chat bots such as HR or companion Chat bots \cite{characterAI} where an end user interacts with a LLM.
Fig. \ref{fig:agent} depicts a common architecture of a GenAI agent that interacts with a user or other agents. It has memory to store prior relevant information and interactions, and other tools to access and send email, browse the web, use datasets and many others. In this infrastructure, the amount of information flow and interaction needs to be governed by policies. But what policies? 

\textbf{What are common policies in the wild?}
We survey available policies in papers and frameworks available. Based on this survey, we grouped policies based on their semantics and showcase where they are applied in the agentic stack. 
Table \ref{tab:policy-categories} compiles a list of representative types of
\textit{policies} we found in the wild.
The list is quite diverse and we broadly classify them in:

\begin{enumerate}
    \item \textit{Regulatory policies} such as GDPR \cite{gdpr}, HIPAA \cite{hippa1996} that are legal frameworks that oftentimes require the enforcement of multiple rules. 

    \item \textit{Content and behavioral policies} that dictate the content that can be exchanged and the types of interactions a machine learning model can have.
    For example private identifiable information (PII) policies mostly require the removal of private information from prompts and assistant answers. Other policies in this category focus on detecting and stopping harms such as hateful speech, misinformation, bias, among others.
    
    \item \textit{Business compliance and logic} where a variety of rules specific to the application are defined. In this category, we see policies that dictate when a person may change their flight for free \cite{yao2024tau}, or when someone is entitled to file an insurance claim for reimbursement \cite{kant2025stanford-law,contracts:stanford:logic}. These are frequently embedded in prompts or application logic. 

    \item \textit{Traditional Cybersecurity Policies} such as identity management, access control and specification of policies to filter out content or stop the access to different tools such as datasets, email, or Internet. 
    In this category we see policies that specify if a tool should be invoked or not by for example scanning file content, looking for vulnerable software, etc. 

    \item  \textit{System Policies} where an agent or application should comply with a service level agreements (SLAs), or other deployment requirements \cite{oscal-compass}.

    \item  \textit{Policies related to the model specification} commonly used to define how LLMs will answer sensitive questions such as \cite{bai2022constitutional,baracaldo:granite:policy,helff2024llavaguard,gemma-model-alginment}.
    In this category, the word policy is also used by the reinforcement learning community to mean the rules an agent uses to receive rewards. Interestingly this may contain the rules themselves or a machine learning model that is assumed to have already learned the right set of rules. 

\end{enumerate}

In the rest of this paper, we refer to \textit{policy} as a set of requirements that once specified can be enforced by a machine.
This means that when we use the word policy, we refer to all the policies listed above except for the ``regulatory policies" which require decomposition and cannot \textit{currently} be directly enforced by a machine. 

\begin{table}[!htbp]
\centering
\begin{adjustbox}{max width=\textwidth}
\begin{tabular}{@{}p{3.5cm}p{9cm}@{}}
\toprule
\textbf{Policy Category} & \textbf{Examples} \\
\midrule
Regulatory Policies & GDPR \newline EU AI Act \newline HIPAA \newline AI Transparency Act \newline Right to be forgotten \\
\addlinespace
Content \& Behavioral Policies & Principle-based harm detectors (e.g., hate, abuse, profanity) \cite{llamaguard,padhi2024graniteguardian}   \newline Mitigate sycophantic interactions \newline Guided conversation  \cite{rebedea2023nemo} Fact \& hallucination control
\newline Alcohol consumption \cite{baracaldo:granite:policy}
\newline Child safety (anthropomorphic, interaction and relational cues \cite{iRaiseGuardrails}, prevent sexual abuse \cite{thornai}) \\
\addlinespace
Model Specification & Alignment policies to define model specifications   \cite{baracaldo:granite:policy,helff2024llavaguard,bai2022constitutional,anthropic:new:constitution:2026,wang2024map} \\
\addlinespace
Business Compliance \& Logic & Punctual application specific logic \cite{yao2024tau,li2025agentorca,huang2025crmarena,kant2025stanford-law} \newline
Business conduct guidelines \cite{achintalwar2024alignment} \newline Don't discuss competitors \newline Enforce business constraints (balance $>$ 0) \newline Ensure app remains on topic \newline Application workflow \cite{itbench-trajectories-2025} \newline Use polite business like tone\\
\addlinespace
Traditional Cyber Security Policies  & Identity management \cite{context-forge,vertexai,cisco-projectGuard,guardrails-ai} \newline Traditional access control \cite{xiang:guardagent,guardrails-ai,context-forge,vertexai,opa-rego} \newline Malware detection \newline MCP file scanners, URL verification \cite{context-forge}   \newline Data flow policies \cite{wang2026:dual:graph:policy:tianYing} \newline Apply secret detectors before storing data \newline Filter private information and competitor names \cite{mellea-hooks} \\
\addlinespace
System Policies & Ensure service level agreements \cite{SLA2026mess+} \newline Container deployment \cite{oscal-compass} \newline Retry twice before scaling out deployment \\
\bottomrule
\end{tabular}
\end{adjustbox}
\caption{Policy means different things to different audiences. 
We compile a diverse set of policy categories that we found in the wild.
GenAI applications typically need to enforce multiple policies simultaneously.
}
\label{tab:policy-categories}
\end{table}

\textbf{What are the targets of GenAI policies?}
Policies apply to a variety of components of the GenAI application as shown in Table \ref{tb:policy-enforcement-places}.
We classify them into \textit{single-turn}, \textit{multi-turn}, \textit{tools}, \textit{trace-based}, \textit{harness} and \textit{memory}. 
The first two columns show if a policy targets single and multi-turn interactions with models.
As it can be seen, policies may apply to single prompt-answer interactions; a great number of guardrails are built to regulate those interactions. We also see that business compliance and logic policies tend to apply to the prompt-answer space. 
Tool interaction is also subject to a variety of policies mostly traditional cybersecurity ones.
Other policies relate to traces of the agentic application that dictate how the application should behave (adequate order of operations). Policies may also need to be enforced at the harness level, for example, a container needs to be deployed only in a particular geography or data can only be stored in certain data centers.
Finally, in the last column, we note that interestingly, policies related to \textit{memory} mostly include contractual stipulations and restricting storage of private data.  

\textbf{What are the enforcement mechanisms?}
The enforcement mechanisms of these policies are quite diverse.
At the same time, GenAI applications need to be governed by one or more policies that need to be enforced in different places in the stack.
To reason about existing approaches and their suitability, in the next section, we propose a framework that helps design applications in a safer fashion.

\begin{table}[H]
\centering
\begin{adjustbox}{max width=\textwidth}
\begin{tabular}{ | p{4cm} | c | c | p{1.2cm} | c | c|c|} \hline

                      & Single-turn        & Multi-turn   & Tools  & Trace-based & Harness &   Memory \\ \hline
  PII Detectors        &  \checkmark        &              & \checkmark       &  &  &\\ \hline
  Harm detectors, fact checking, hallucinations \cite{rebedea2023nemo,llamaguard,padhi2024graniteguardian}      &  \checkmark        &              &                  &   &  & \\ \hline
  Sycophancy           &  \checkmark        & \checkmark   &           &       &   & \\ \hline
  Filter competitors names &  \checkmark        &              &        &          &  &\\ \hline
  Secret detectors     &  \checkmark        &              & \checkmark  &     &    & \\ \hline
 Code Safety          &  \checkmark        &              & \checkmark    &   & & \\ \hline
  OPA regex            &                    &              & \checkmark   &  &  & \\ \hline
   URL verification     &                    &              & \checkmark    &  & & \\ \hline
  MCP file scanners \cite{context-forge}   &                    &           &   & \checkmark  &     & \\ \hline
  Data flow policies \cite{wang2026:dual:graph:policy:tianYing}   &        &            &              & \checkmark       & \checkmark  & \\ \hline
  Traditional access control \cite{xiang:guardagent,guardrails-ai,context-forge,vertexai}&                    &              & \checkmark    & &  & \\ \hline
  Identity management \cite{context-forge,vertexai,cisco-projectGuard,guardrails-ai} &                    &              &  \checkmark      &  & &  \\ \hline
  Flow-based policies  \cite{wang2026:dual:graph:policy:tianYing}     &          &          &              & \checkmark       &  \checkmark & \\ \hline
  Punctual application specific logic  \cite{yao2024tau,li2025agentorca,huang2025crmarena,kant2025stanford-law} &  \checkmark        &              & \checkmark       & & & \\ \hline

 Business conduct guidelines \cite{achintalwar2024alignment} & \checkmark  &    \checkmark & \checkmark  &     &  \checkmark  & \\ \hline
 General formatting requirements & \checkmark  &   &   &  &  &\\ \hline
 Model alignment policy \cite{helff2024llavaguard,bai2022constitutional,wang2024map} & \checkmark  & \checkmark  &   &  & & \\ \hline

Deployment \cite{oscal-compass}  & & & & & \checkmark & \\ \hline

Agentic memory \cite{memory-anthropic,memory-openai,asif2026lcguard:kvcache:privacy}  & & & & & & \checkmark\\ \hline 

\end{tabular}
\end{adjustbox}
\caption{Targets of GenAI policies}
\label{tb:policy-enforcement-places}
\end{table}

\section{Proposed Approach to Dissect Policies} \label{sec:approach-to-dissect-policies}

We now propose a methodology to dissect policies and their respective enforcement mechanisms to understand how they differ and what are their gaps with the objective of reducing the noise related to inappropriate enforcement proposals.

We dissect current approaches by analyzing the following dimensions: 
\begin{enumerate}
    \item \textbf{Policy Definition} can vary between \textit{strict} and \textit{fuzzy}.
    Strict policies are frequently specified using formal languages such as XACML. With the capability improvement of LLMs, it is possible to specify policies using natural language which oftentimes is less precise.

    \item \textbf{Mechanism Enforceability} 
    Policy enforcement mechanisms can be characterized based on the guarantee enforceability that they provide, which may fall in a spectrum between \textit{soft} and \textit{hard} enforceability. 
    In one extreme, \textit{hard} mechanisms are \textit{deterministic} and often provide guarantees about their enforcement. Traditional cybersecurity solutions such as enforcing identity management, role-based access control, among others have been designed to have hard enforceability. 
    In the other side of the spectrum, \textit{soft} enforceability mechanisms are stochastic in nature and rely on machine learning models or LLMs to enforce a policy.
    Given this stochasticity, there is no guarantee of success.
    
    \item \textbf{Stack Enforcement}
    This dimension refers to the place where the policy is enforced within the GenAI application.
    Common places include: the LLM itself (e.g., activation steering) or 
    specific places in the application such as MCP server.
    This dimension is also critical for \textit{defense in depth} which at its core requires instantiating multiple mitigation mechanisms.
    
\end{enumerate}

\subsection{Interactions between policy definition and enforceability}
\begin{figure}[htbp]
    \centering
    \begin{subfigure}[b]{0.49\linewidth}
        \centering
        \includegraphics[width=\linewidth]{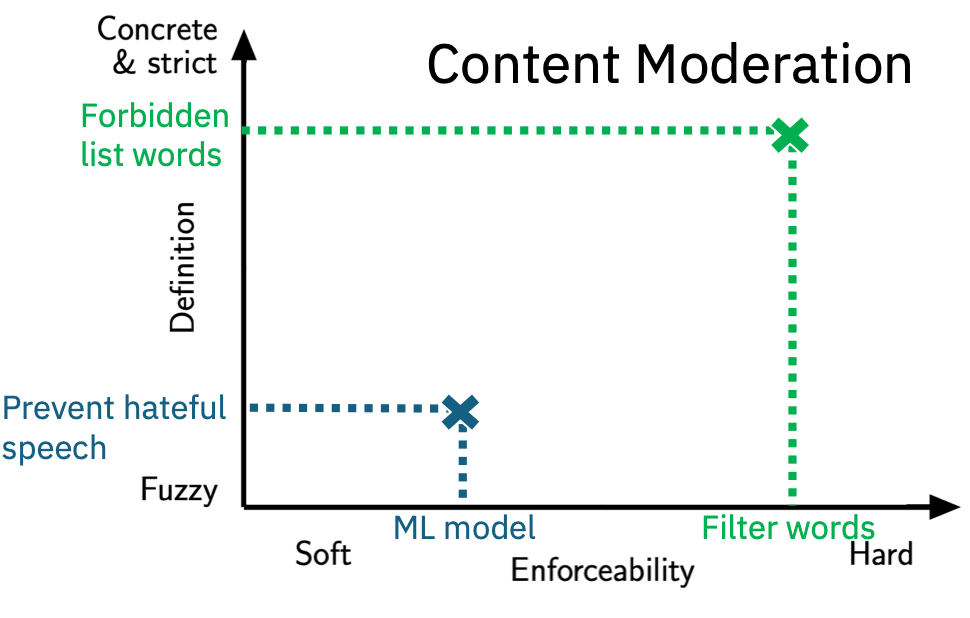}
        \caption{Content Moderation}
        \label{fig:content-moderation}
    \end{subfigure}
    \hfill
    \begin{subfigure}[b]{0.49\linewidth}
        \centering
        \includegraphics[width=\linewidth]{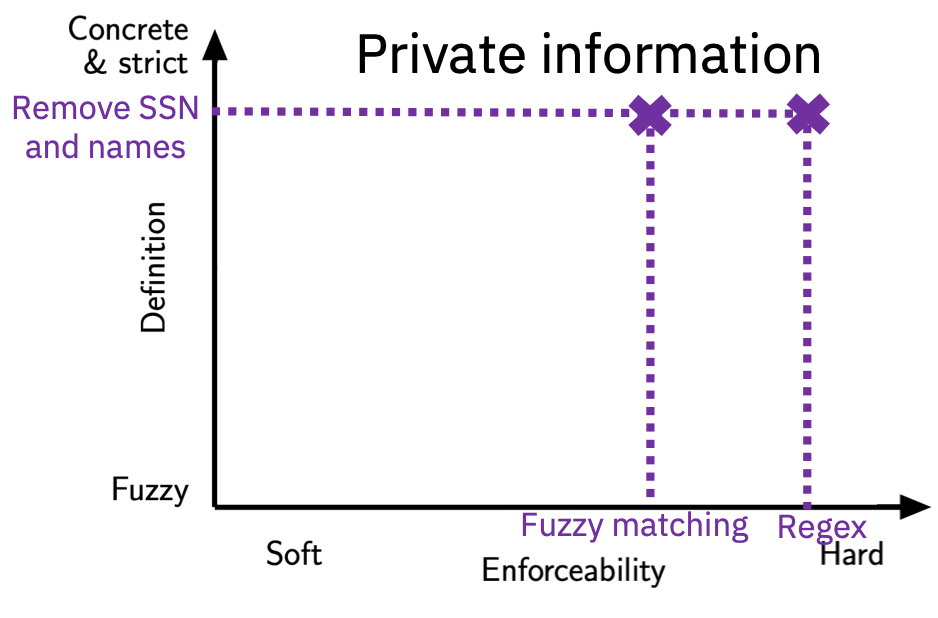}
        \caption{Private Information}
        \label{fig:private-information}
    \end{subfigure}

    \vspace{1em}

    \begin{subfigure}[b]{0.49\linewidth}
        \centering
        \includegraphics[width=\linewidth]{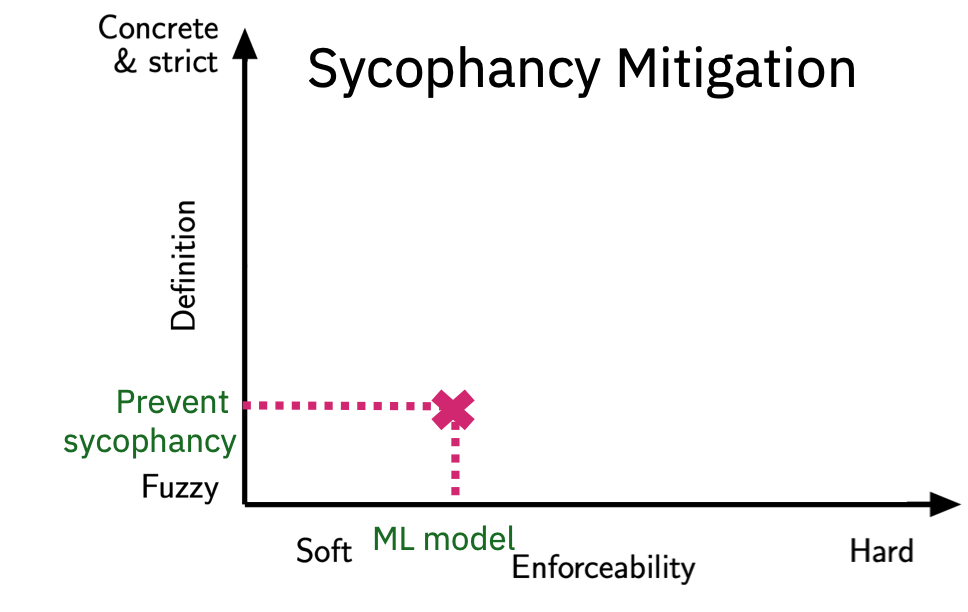}
        \caption{Sycophancy Mitigation}
        \label{fig:sycophancy-mitigation}
    \end{subfigure}
    \hfill
    \begin{subfigure}[b]{0.49\linewidth}
        \centering
        \includegraphics[width=\linewidth]{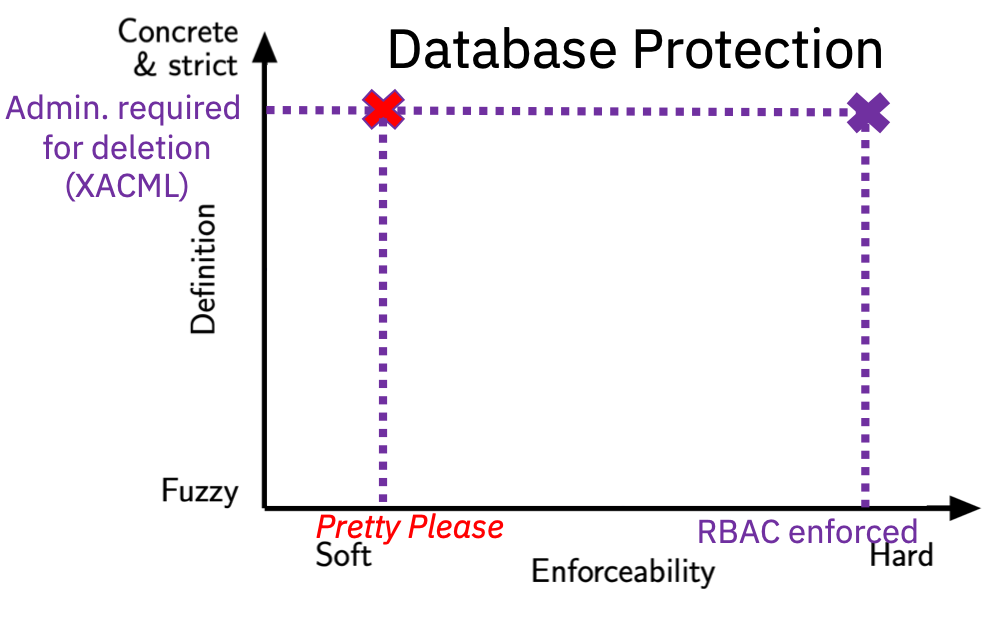}
        \caption{Database Protection}
        \label{fig:database-protection}
    \end{subfigure}

    \caption{Four use cases mapped by definition clarity (fuzzy to concrete) and enforceability (soft to hard).}
    \label{fig:example-policy-interaction}
\end{figure}

We now analyze the interactions between the \textit{policy definition} and the \textit{enforceability} dimensions.
We use a graphical representation with two axis as depicted in 
Fig. \ref{fig:example-policy-interaction}.
In practice, there may be multiple ways in which a policy definition may be enforced.
This flexibility and the newer capabilities of LLMs sometimes lead to undesirable selections that compromise the security, safety and compliance of the system.

\textbf{Understanding the quadrant.}
We define a tuple  $T = \langle \mathcal{P}, \mathcal{E} \rangle$ where its first component $\mathcal{P}$ contains the policy definition, and the second $\mathcal{E}$ the enforcement mechanism used. Fig.~\ref{fig:example-policy-interaction} locates example tuples across different policy domains.
The objective is to locate tuples of policy definition and enforcement mechanisms in \textit{general} regions rather than generating strict orders among tuples.

As a first example consider \textit{Content Moderation} (Fig. \ref{fig:content-moderation}) shows two different policy definitions one very concrete: ``Forbidden list words" which can be enforced deterministically (hard enforcement), vs. a more fuzzy definition ``Prevent hateful speech" which can be enforced using a machine learning model guardrail trained for that purpose resulting in stochastic enforcement e.g., \cite{padhi2024graniteguardian,guardrails-ai,llamaguard}.
Both these policy enforcement mechanisms are traditional and adequate. In some cases using both of them simultaneously is suitable to capture policy violations.

In \textit{Private Information} (Fig. \ref{fig:private-information}) ``Remove SSN and names" is a concrete definition that can be enforced with hard controls like regex matching. This enforcement can be combined with a more stochastic policy enforcement such as fuzzy matching \cite{cayrol1982fuzzy-matching} to filter private data.

Another policy ``Prevent sycophancy" (Fig. \ref{fig:sycophancy-mitigation} ) has inherently fuzzy definition and requires soft enforcement.
The concept of sycophancy in itself is fuzzy \cite{cheng2025elephant} with the research community still studying the phenomenon.
Current enforcements mechanisms to mitigate sycophancy include activation steering \cite{cheng2025elephant,panickssery2023steeringPycophancy}, a technique that changes the LLMs behavior stochastically at runtime.
There is no enforcement mechanism that is deterministic and offers provable results; yet not mitigating this risk can lead to catastrophic life-death situations \cite{characterAIteen}.
In this case, the enforcement is soft, but given the type of policy definition, it is the best that can be done.

Let's examine policies related to database protection (Fig. \ref{fig:database-protection}). The cybersecurity community enforces policies that restrict access to resources such as dataset by specifying the policy using formal language and using enforcement mechanisms that are carefully placed in the application stack to have provable guarantees of enforcement
(Section \ref{sec:sota} provides more details about these solutions and their properties).
A new set of enforcement mechanisms have taken GenAI applications by storm.
We call them \textit{Pretty Please policy enforcement} and have very different properties compared to traditional cybersecurity enforcement mechanisms. 

\textbf{The rise of the \textit{Pretty Please} Policy Enforcement.}
With the increase of instruction following capabilities of LLMs, policy definitions have changed dramatically enabling users to specify policies in natural language -- something that was not possible before.
What's more interesting is the fact that any policy written in natural language can be enforced (or at least give the impression of being enforced) asking the LLM to adhere to the desired policy. We call this \textit{Pretty Please policy enforcement}.
This type of enforcement is typically done through system prompt (where main directives of operation are typically given to the LLM) and \textit{in-context learning} \cite{brown2020languageICL} (where multiple examples of \textit{good} and \textit{bad} behavior are provided in a prompt). Hence, there is no enforcement guarantee.

A real example of a \textit{Pretty Please} policy can be seen in Fig. \ref{fig:pretty-please-router} where the prompt to define a router includes information about the task at hand and the policy by which such router should be governed
\textit{``Strictly adhere to the following rules: never shared personal data, always verify claims, and use markdown policy"} \cite{screenshot:pretty:please}. In this setting, the enforcement is done by the router which is implemented as an LLM. This is a \textit{Pretty Please} with \textit{no} enforcement guarantee.

\begin{figure}
    \centering
    \begin{tcolorbox}[
        colback=gray!5,
        colframe=gray!75,
        width=0.85\linewidth,
        arc=2mm,
        boxrule=0.5pt,
        title={\textbf{Example: Router/Constitution Prompt}},
        fonttitle=\sffamily
    ]
    \textit{``You are a master productivity router. Your job is to analyze the user's request, delegate it to the appropriate sub-agent (Research, Writing, or Coding), and synthesize their final response.}
    \vspace{0.1em}
    \begin{enumerate}[leftmargin=*, itemsep=0.2em]
        \item \textbf{Governance:} Strictly adhere to the following rules: never share personal data, always verify claims, and use markdown formatting.
        \item \textbf{Conditional Logic:} If the user requests code, route to the Coding Agent. If they request long-form text, route to the Writing Agent.
        \item \textbf{Output:} Provide a brief summary of which agent you chose and why, followed by the final deliverable.''
    \end{enumerate}
    \end{tcolorbox}
    \caption{Example prompt to guide a router agent. Governance constraints are specified as part of the prompt. While providing these instructions may \textit{help} the LLM generate suitable plans, they should be taken as a \textit{Pretty Please} request. There is no guarantee they will be fulfilled.}
    \label{fig:pretty-please-router}
\end{figure}

The deletion incident of the production database discussed in the introduction \cite{deletion-insident} (Fig. \ref{fig:incidents}) was caused by the use of a soft \textit{Pretty Please policy} instead of a hard, well-establish policy enforcement mechanism.
In this case, good security practices require having access control policies written in languages such as XACML or OPA \cite{opa-rego} and enforced by carefully implemented engines injected in enforcement points that cannot be circumvented to achieve as much as possible deterministic enforcement.
Fig. \ref{fig:database-protection} shows the difference between these enforcement in the analysis quadrant. The use of \textit{Pretty Please} policy enforcement is not isolated, for example, it has been used in \cite{xiang:guardagent,zeng2024airbench,yao2024tau,li2025agentorca,huang2025crmarena}.

What is both deeply interesting and troubling about the \textit{Pretty Please} enforcement is that \textit{any} type of policy definition can be \textit{attempted} to be enforced that way: all that's needed is a prompt.
Easiness of specification and deployment seems to be the reason for adoption despite lack of enforceability.
Let's examine in more detail the risks associated with \textit{Pretty Please} policy enforcement.

\textbf{The perils of the \textit{Pretty Please} Policy Enforcement.}
This enforcement mechanism is problematic because it relies solely on the LLM, a highly stochastic system, to enforce the policy.
It is equivalent to ask \textit{Pretty Please don't delete my database} - it may or may not work.

In addition, enforcement through LLMs is subject to
a variety of undesirable behaviors: 
\begin{itemize}
    \item \textit{Non-adversarial settings:} even under normal circumstances LLMs have been shown to 
    \textit{1)} exhibit unpredictable stochastic errors \cite{jha2026agent-metldowns-vitaly} where, for example, a benign error, like a missing file or a 404 page, may trigger the agent to follow undesirable and sometimes dangerous actions such as conducting unauthorized reconnaissance or subverting access control,
    \textit{2)}~not follow system prompts adequately \cite{mccauley2026HIBenchmarkJasonHiddenLayer},
    \textit{3)} suffer from \textit{reward hacking} \cite{amodei2016concreteRisks} where the model attempts to complete a task at all costs to get its final reward (task marked as successfully completed) even if the actions are unethical or otherwise undesirable and also, 
    \textit{4)} collude with other agents in unpredictable ways \cite{metr-2026-openai-hugging-face-incident-investigation}. 
    
   \item \textit{Adversarial vulnerabilities:} LLMs are subject to jailbreaks, direct and indirect prompt injections \cite{owasp-llm,liu2024formalizing} that make them a target to adversaries motivated to circumvent any \textit{Pretty Please} policy as demonstrated by FragFuse \cite{rao2026fragfuse-usenix-boli}.
\end{itemize}

For all these reasons, relying on \textit{Pretty Please} policy \textit{enforcement} is not recommended when there is a route to enforce through hard mechanisms. 
To be clear, there is nothing wrong in providing a \textit{Pretty Please} \textit{description} to \textit{guide} the LLM in the selection of suitable agentic plans; what we are highlighting is that a description of desired behavior should not be confused with \textit{adequate enforcement}.

\begin{theorem}
    Don't rely on \textit{Pretty Please} prompts to enforce policies, they are meant to guide the LLM behavior (without guarantee), but are not suitable to ensure the policy is enforced.
\end{theorem}

\begin{figure}
    \centering
    \includegraphics[width=0.65\linewidth]{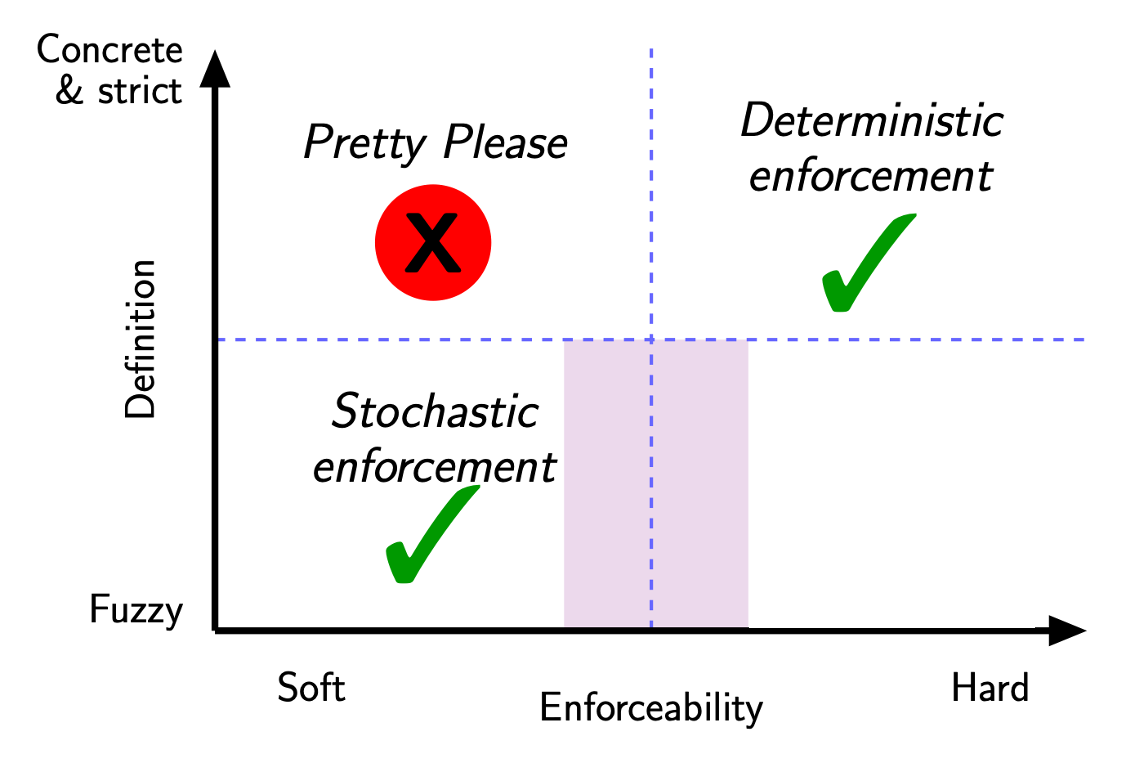}
    \caption{Recommendation: Not all policies are the same. They need different enforcement mechanisms depending on the policy specification.}
    \label{fig:takeaway-quadrants}
\end{figure}

In Fig. \ref{fig:takeaway-quadrants},
we split the space of interaction between policy definition and policy enforceability in five regions according to their suitability:

\begin{enumerate}
    \item \textit{Deterministic enforcement region:} This is a desirable region. In this region lie policies where a concrete definition is paired with a hard deterministic enforcement. 
    
    \item \textit{Stochastic enforcement region:} This is a desirable region for fuzzy policies that can only be enforced with stochastic methods like the sycophancy example.
    
    \item \textit{The Pretty Please region:} 
    This is an undesirable region where despite the existence of a concrete policy definition and deterministic enforcement mechanisms, a \textit{Pretty Please} enforcement is selected. Deployments should aim to select enforcement mechanisms that are hard.

    \item \textit{Semi-stochastic region (purple):}
    In this region, we see a rise of solutions that try to reduce stochasticity of purely using generative models by dividing complex instructions into atomically in a way that the LLM work is closely monitored, verified and re-done if necessary. Example approaches include Mellea where policies can be enforced through atomic functions \cite{mellea}. 
    
\end{enumerate}

\begin{theorem}
Prefer hard policy mechanisms when the policy definition supports them, following Fig. \ref{fig:takeaway-quadrants} recommendations.
\end{theorem}

\subsection{The Stack Enforcement and Defense in Depth}
The third dimension we analyze as part of the proposed methodology is the \textit{stack enforcement}.

\subsubsection{Defense in Depth}
\begin{figure}[htbp]
    \centering
    \begin{subfigure}[b]{0.5\linewidth}
        \centering
        \includegraphics[width=\linewidth]{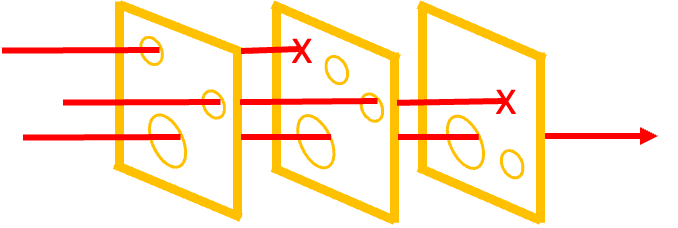}
        \caption{Defense in Depth: Swiss Cheese Model}
        \label{fig:swiss-cheese-model}
    \end{subfigure}
    \hfill
    \begin{subfigure}[b]{0.75\linewidth}
        \centering
        \includegraphics[width=\linewidth]{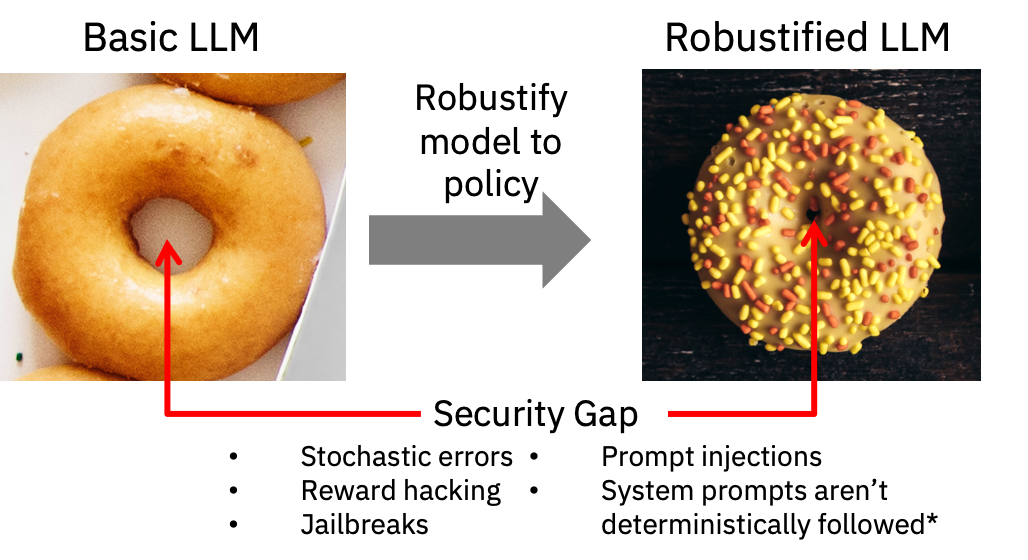}
        \caption{Donut-LLM analogy}
        \label{fig:analogy-donut}
    \end{subfigure}

    \caption{Analogy: If an LLM was \textit{a donut}, we could see its security gap (Fig. \ref{fig:analogy-donut}). The \textit{Pretty Please} prompts do not close that gap. Methods to robustify the model such as alignment, fine tuning, and unlearning may help reduce the gap -- resulting in a sprinkle donut. To this date, no method guarantees the gap will be fully closed. A \textit{donut} is a single layer of defense in the Swiss Cheese model.}
    \label{fig:analogy}
\end{figure}

An additional important distinction between enforcement mechanisms is where in the stack the enforcement takes place.
 Defense in depth is an important factor in cybersecurity and governance with the Swiss cheese model \cite{swiss-cheese-model} (Fig. \ref{fig:swiss-cheese-model}) exemplifying how to ensure multiple layers of defenses are orchestrated to provide defense in depth.

During our tour of the wild, we uncovered a common pitfall that we explain through an analogy:
some practitioners are confusing a \textit{donut} (term coined in this paper) with the Swiss cheese model (Fig. \ref{fig:analogy}).

At the core of any GenAI application there is an LLM. If the LLM was a donut, we could see its security hole consisting of  
stochastic errors,
jailbreaks,
prompt injections and lack of reliable prompt following capabilities \cite{jha2026agent-metldowns-vitaly,mccauley2026HIBenchmarkJasonHiddenLayer}. 
\textit{Pretty Please} policies do not close this security gap, thus using them does not reduce risk in a reliable way.

To close the security gaps of LLMs, techniques such as alignment, unlearning and fine tuning have been developed. These techniques manipulate the LLMs to increase its reliability for different tasks. When they are used to improve policy compliance, they can indeed reduce the security gap of the model. In particular, when these model adaptations are tailored by policies or policy-driven red teaming. Going back to the analogy, the adapted model is a sprinkle donut with a smaller security gap. However, it is important to note that the LLM (donut) will always have some stochastic component and a security gap that, to this date, cannot be fully closed.
The \textit{sprinkled donut} is more robust than the \textit{generic} donut: always prefer a sprinkle donut. However, a donut with or without sprinkles remains a single layer of defense. Whether there is a \textit{Pretty Please} policy in place or not, this fact does not change.  

True defense in depth requires more than using \textit{Pretty Please} prompts, or robustifying an LLM (donut with sprinkles). It requires having multiple components orchestrated across the application. Avoid reusing the same model to guard itself to mitigate its resulting actions suffering from reward hacking, jailbreaks or injection prompts.

\begin{theorem}
True defense in depth requires multiple layers of protection. A single model, even if it is robustified, or if it has been given \textit{Pretty Please} prompts, should be considered as a single layer of protection. It needs to be deployed with further safeguards. 
\end{theorem}

\subsubsection{Using Policy Enforcement Points}
The number of places where policy needs to be enforced in GenAI application is substantial (Table \ref{tb:policy-enforcement-places}).
The design principles from traditional cybersecurity are still valid in this GenAI era (see Section \ref{sec:policy-enforcement-principles} for a detail review). In particular,
separating policy definition, policy enforcement points through \textit{hooks} that can integrate diverse enforcement mechanisms are some of the relevant design principles.
A \textit{hook} (as defined by Linux Security Modules \cite{wright2002linux-security-modules})
should be placed in a spot in the application where policy enforcement is required. As shown in Fig. \ref{fig:agent} and Table \ref{tb:policy-enforcement-places}, there are multiple places in the stack where policy enforcement is needed. Ensuring hooks are placed in the right spots and tied to the right enforcement mechanism is important.
Adding application hooks and linking them to hard enforcement mechanisms to reliably and deterministically verify certain actions, such as authentication and verification of access tokens is a must, can substantially improve the security of the system.
For this purpose, architecting the right abstractions in the application is important, for example, using MCP \cite{mcp} to help determine when an agent makes certain tool calls and then enforcing policies for example using MCP Context Forge \cite{mcp-context-forge}.
Recent work \cite{cosai:zerotrust,mellea-hooks}, shows how hooks have been inserted to enforce access control, identity management policies and various other policies. 
More details about these design principles and some systems implementing them can be found in Section \ref{sec:sota}. 

\begin{theorem}
Use well-defined policy enforcement points and maintain a clear policy definition.
\end{theorem}

\subsection{Enforcing Multiple Policies}
GenAI application may require the enforcement of a variety of policies (Fig. \ref{fig:policy-clusters}) which need to be enforced in different places in the stack: model, tools, prompts and memory.
Additional complexity arises from the fact that the enforcement mechanisms are quite diverse as shown in Table \ref{tb:policy-enforcement-places} and Fig. \ref{fig:multiple-policies}. 
We identify the following challenges:

\begin{figure}
    \centering
    \includegraphics[width=0.6\linewidth]{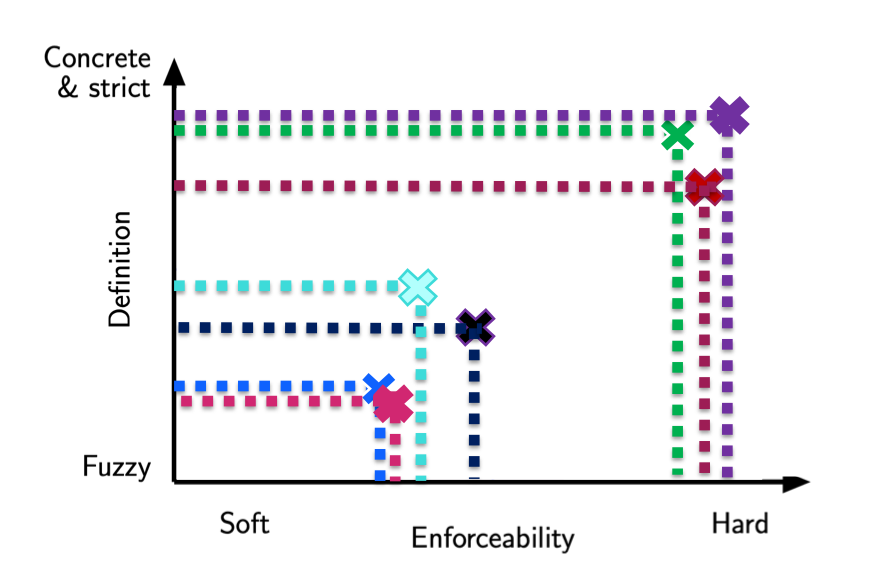}
    \caption{GenAI applications may need to comply with a variety of policies located in different regions of the quadrant imposing challenges to compliance}
    \label{fig:multiple-policies}
\end{figure}

\textbf{Lack of a single control plane}
Ideally all policies could be visible from a single place. Currently, diverse mechanisms exist to enforce specific policies, e.g. OPA/REGO, Nemo Guardrails, among others. But they all limit themselves to limited types of policy. Using hooks for \textit{any} enforcement is a good design principle to achieve this objective.

\textbf{Log and agree on error definition “common errors”}
GenAI applications are bound to fail from time to time. With the increase of multi-agent applications, identifying where issues occur is important for compliance and recovery.
Logging interactions is a must, and understanding what to log and communicate across agents is also very important.
For example, in http, the error 404 means ``Not Found error" which means the specific page or file requested does not exist.
A recent approach \cite{baracaldo:granite:policy},
proposes defining policy definition 
that includes exception codes that should be triggered when failures occur.
Implementing such type of policy enforcement and exception raising would improve compliance. 

\begin{theorem}
Having a central control panel, capable of logging well specified errors, with ideally \textit{universal} error codes can facilitate compliance.
\end{theorem}

\textbf{Conflict resolution}
When there are more than one policy,
there may be conflicts \cite{shafiq2005coflictsrbac,joshi2004access,cuppens2007conflictresolution,huynh2019sgac}. 
For example, one policy may require restricting an agent from sending emails, while another one may require sending a message to an emergency contact when there is a self-harm emergency. Ensuring policy resolution is congruent with real objectives remains an open issues. The great majority of solutions, e.g., \cite{rebedea2023nemo}, rely on content similarity of user request to define the right policy to apply.

\begin{theorem}
In practice, multiple policies need to be enforced by a single GenAI application. 
The policies have \textit{fuzzy} and \textit{concrete definitions} and require both \textit{soft} and \textit{hard} enforcement mechanisms.
They need to co-exist and be enforced together to prevent undesirable side effects due to conflicting objectives and obligations.
\end{theorem}

\section{Related work} \label{sec:sota}
Enforcing policies in GenAI applications and model is a popular topic.
In this section, we present in more detail representative approaches clustering them based on the field they are proposed starting with cybersecurity.

\subsection{Traditional Cybersecurity Policy Enforcement} \label{sec:policy-enforcement-principles}
Cybersecurity has studied extensively policy enforcement for traditional (non-GenAI) applications. Most of this work focuses on enforcing access control policies such as role-based access control (RBAC) \cite{ferraiolo2001rbac}, attribute-based policies (ABAC) \cite{hu2015abac},
 flow related access control such as BIBA's \cite{biba1977integrity} and Bell-Lapadula \cite{bell-lapadula1973secure}.
A wide range of application domains such as healthcare, information systems also require enforcing \textit{obligations} \cite{irwin2006obligations,pearson2009accountability,hippa1996,li2012pobligation,baracaldo2016tackling}, which dictate what actions need to be applied before and/or after the evaluation of a policy within a specify amount of time. Obligations are combined with a variety of access control models.

The security community has also focus widely on enforceability and composability of policies.
Access control models are frequently modeled as state machines and policies are described using grammars and policy languages to ensure policy enforceability. Based on these clear defined abstractions, it is possible to analyze composability of policies.
Composability of multiple policies may lead to conflicts  \cite{shafiq2005coflictsrbac,joshi2004access,cuppens2007conflictresolution,huynh2019sgac}.
Approaches in this area aim to determine conflicts of policies written in grammars that can be reasoned upon.
Conflicts are resolved using rules that specify what policy should prevail; in healthcare, conflict resolution may also include  \textit{break-the-glass} policies where a system may violate \textit{normal} policy under very specific circumstances \cite{brucker2009breaktheglass}. For example, denying an access to a healthcare chart may result in death if the patient is in ER. A break-the-glass policy states that a doctor in ER taking care of the patient may access his chart even if she is not the primary physician.
Additional work has pointed out the necessity of ensuring enforceability of obligations and finding potential conflicts that may prevent an agent or user from completing one such obligation \cite{baracaldo2016tackling,chowdhury2012cascadingobligations,baracaldo2013beyond}, for a user process may not be able to complete an obligation if it lacks of privileges to carry out the require action. 
Well-defined grammars are in sharp contrast with soft policies (written in natural language and LLM-based detectors), we have found in the wild, e.g., \cite{xiang:guardagent,zeng2024airbench,yao2024tau,li2025agentorca,huang2025crmarena}.  

\begin{figure}[h!]
    \centering
    \includegraphics[width=0.9\columnwidth]{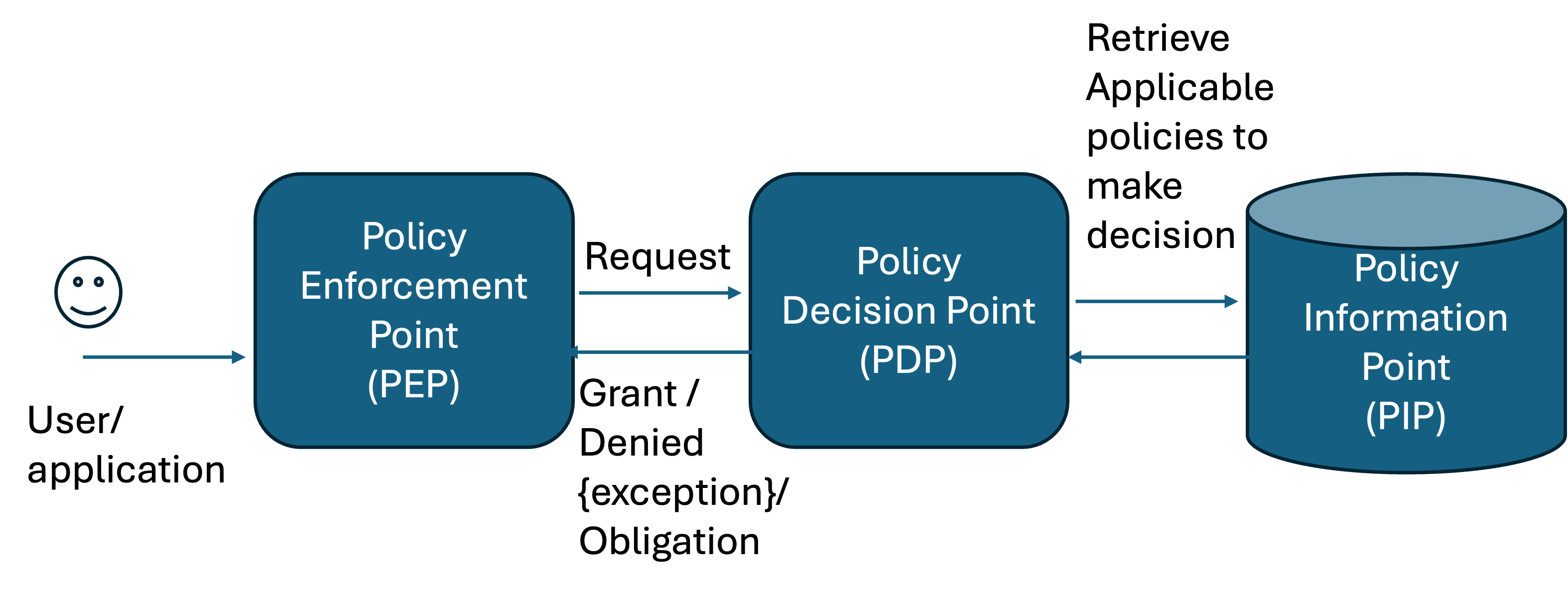}
    \caption{  
    Policy enforcement for traditional access control applications. The design enables modularization and improves enforceability especially for distributed applications. Having an independent PIP facilitates updating policies dynamically. This design can also be used for GenAI applications.}
    \label{fig:traditional-enforcement}
\end{figure}

Good policy enforcement designs have also been a core component of cybersecurity.
Among the aspects that have been widely discuss are the appropriate design of applications that need to ensure enforceability of policies with minimum modification of application code to ensure easy maintenance.
In particular, designing applications ensuring \textit{policy enforcement point} (PEP), \textit{policy decision point} (PDP) and \textit{policy information point} (PIP) provides flexibility and enforceability in a modular way.
Figure \ref{fig:traditional-enforcement} shows this architecture. Policy specifications are stored in the PIP ensuring they can be updated without modifying the application or other components. The PEP serves as an execution monitor \cite{schneider2000enforceable} ensuring interactions are intercepted and calling the PDP, which decides if there are any policy violations by querying the PIP for relevant policies. In this way, it is possible to ensure policies are maintained separately to the application's logic. They can also be one or more PIPs to ensure reliability. Traditionally, the PDP returns to the PEP a \textit{grant/deny} that it is then enforced by the PEP by blocking access or granting access to the particular resource. In some special cases when \textit{obligations} exist, the PDP returns the obligation(s) that needs to be fulfilled, and the PEP is in charge of completing the obligation.
In cases where there are conflicts in the PIP, the PDP finds them and decides what to do with such conflict. Addressing conflicts in general requires reasoning among policies written in well established grammars.
In contrast, a lot of the frameworks that inspect natural language violations in GenAI address conflicts through embedding similarity, e.g., \cite{rebedea2023nemo}.

Linux Security Module (LSM) \cite{wright2002linux-security-modules} is a framework that supports a variety of security access control models in Linux Kernel through \textit{hooks} and follows similar design principles separating PIP, PDP and PEP.
In LSM, a hook is a function with a well defined signature placed at critical security-relevant points in code of the kernel --effectively making it a PEP. A \textit{module} can be \textit{plugged} into the \textit{hooks} so that it is automatically called at these key points, and the result is used for enforcement, such as blocking the execution of the process.
This design ensures modules can change without requiring modifications to the kernel itself.
This design ensures the GenAI application does not need to be modified. 
Systems embracing hook architecture include \cite{mellea-hooks}.
An example design with this zero-trust in mind was presented in \cite{cosai:zerotrust}.

The GenAI area has brought with it a very different reality, where policies are broader and generally written in natural language. This is completely opposite to the very rigid and well-defined grammars used for traditional access control. We now overview additional types of policies and their enforcement.

\subsection{Policies for System, Business Compliance and Logic}
Multiple approaches in this bucket propose using natural language policies and
transforming them into intermediate representations \cite{zwerdling2025towards,contracts:stanford:logic,achintalwar2024alignment,kant2025stanford-law}.
In \cite{zwerdling2025towards}, a method that transforms policy documents into verifiable code associated with tool usage is provided. At runtime, these rules are verified to check compliance before the agent performs an action.
This work limits itself to tool invocation verification.
A similar work in \cite{contracts:stanford:logic} uses a LLM to transform into Prolog text-based policies written in natural language that dictate when a health insurance claim is valid or not. At runtime, the claim is fed to the LLM and the policies are evaluated in Prolog. A similar approach for law contracts was presented \cite{kant2025stanford-law}.
Another approach \cite{achintalwar2024alignment} proposes aligning the model to answer questions according to multiple policies. This method mainly deals with question-answer type of interactions.
An example approach that deals with system related policies is OSCAL (Open Security Controls Assessment Language) created by NIST \cite{oscal-compass}. It aims to automate cybersecurity compliance by providing standardized, machine readable formats.
None of these methods have been designed to comply with the multi-stakeholder policies.

Multiple approaches evaluate policy adherence by appending it directly to the prompt before querying the LLM \cite{yao2024tau,li2025agentorca,huang2025crmarena} falling into the \textit{Pretty Please} enforcement.
Li et al. \cite{li2025agentorca} propose a SOPBench, a benchmark to measure compliance of
standard operating procedures (SOPs), e.g.,
``To schedule a driving test, please verify that the user is at least 16 years AND has passed the knowledge test."
SOPs described in natural language are transform into source code (oracle code), then LLMs are queried to evaluate if agents’ tool-calling
trajectories follow the oracle code trajectories.
The evaluation includes the SOP instruction as part of the prompt received by the LLM.
Yao et al. \cite{yao2024tau}, propose $\tau$-bench 
a popular benchmark to measure policy compliance for airline use cases. 
The evaluations are performed by adding the domain policy as system prompt and user's instructions as user prompt, respectively.
A more elaborated benchmark \textit{CRMArena} measures Customer Relationship Management (CRM) that mimic Sales force CRM  \cite{huang2025crmarena}.
Evaluation is performed by feeding a long prompt containing all the description of requirements and policy to the LLM or agent.
Some of these works use \textit{policy} to refer to \textit{application logic rules} or requirements.
As discussed before, relying in these \textit{Pretty Please} enforcement may lead to undesirable behavior.

\subsection{Enforcing Policy by Adapting Models or Their Behavior}
At their core, GenAI applications rely on LLM models to generate content
and for that reason a variety of techniques have been proposed to modify models weights according to policy.
These techniques include model
alignment \cite{christiano2017rlhf,shao2024grpo,wang2024map,ding2025improvedSFT,bai2022constitutional} and unlearning \cite{liu2024rethinkinguUlearning,wang2025invariance:unlearning,unlearning:book:2026,wang2025rethinking}.
On one hand, model alignment consists on training the model to \textit{align} it with a policy \cite{baracaldo:granite:policy,helff2024llavaguard}, which is also referred to as a model specification \cite{gemma-model-alginment} or
constitution \cite{bai2022constitutional}.
A variety of optimization methods exist for this 
purpose including RLHF \cite{christiano2017rlhf}, multi-human-value alignment palette (MAP) \cite{wang2024map}, GRPO \cite{shao2024grpo}, SFT \cite{sft-llms-blog,ding2025improvedSFT}, DPO\cite{rafailov2023dpo}, among others.
Unlearning on the other hand,
has been proposed to carefully remove unwanted behavior or data from the model by carefully changing the model once a problem has been detected.
A variety of approaches have also been designed to train LoRA \cite{hu2022lora} and aLoRA adapters \cite{greenewald_activated_2025}, where the latter are activated at runtime via a special token. 

An additional trend is to modify model behavior at runtime by steering the model's activations \cite{cheng2025elephant}. Recently \cite{rozenfeld2026gavel} proposes marrying the policy rules with activation monitoring to ensure policy failures are corrected on the fly at runtime.
These methods are promising and further study on their potential side effects is needed.

Prior work \cite{baracaldo:granite:policy,llamaguard} proposed YAML-based policy specifications to define how the model should answer.
In \cite{baracaldo:granite:policy}, besides specifying what the model can and cannot output, the policy also defines the \textit{exception} that should be triggered when policy gets violated facilitating compliance verification. This type of setup can help understand failures across multiple agents and policies.

\subsection{Stochastic Methods and Efficacy Evaluation}

GuardAgent \cite{xiang:guardagent} aims to enforce policy in agents by generating plans that consider a policy. It relies solely on prompt-based interventions (even for traditional access control policies). 
ShieldAgent \cite{chen2025shieldagent-boli}
uses a probabilistic policy reasoning enforcement module where policy constraints are compiled from documents to generate action-based rule circuits. During setup, each circuit associates an agent action with relevant rules for verification.
At runtime, ShieldAgent \textit{i)}. receives the plan generated by the LLM it guards, 
\textit{ii)} formats it and retrieves similar workflows (from long term memory where what has been labeled as safe interactions have been previously stored)\footnote{This can be problematic if access has been revoked, for example, if the employee is no longer at the company.},
\textit{iii)} uses them as in-context learning samples with the objective of generating a safe plan and \textit{iv)} verifies each rule and makes a final probabilistic inference.
AGrail \cite{luo2025agrail} proposes a risk mitigation framework that
uses multiple LLMs to detect generic \textit{universal} and manually specified risks.
Embedding vectors are used to verify what risks are relevant for a given request.
Conseca \cite{tsai2025consecagoogle} was proposed to secure general purpose agents,
where anticipating all relevant contexts, tasks and proving policies for them may not be feasible. To overcome the hurdle of specifying policies for each potential context manually, Conseca uses an LLM to generate on the fly policies for fine-grained contexts. Their implementation uses hooks and a deterministic policy enforcer; the policy itself, however, is generated based on in-context learning examples and non-deterministic.
AuthGraph uses an authorization graph and a provenance graph to prevent malicious accesses. The construct increases contextual information to make it more difficult to subvert the agent, but still relies on an LLM is used for attribution capabilities.

\textbf{Evaluation.}
FragFuse presents a method to bypass LLM-based access control mechanisms \cite{rao2026fragfuse-usenix-boli} by manipulating the instruction that is sent to the LLM-based access control module. The method modifies long term memory by splitting the undesirable instruction into chunks that are mix with other benign content. Four agents were evaluated using LLM-based access control that consisted of a \textit{Pretty Please policy} injected in the prompts which were unsurprisingly broken.
FragFuse also tested GuardAgent, AGrail and ShieldAgent as enforcement mechanisms.
All these methods were broken by the attack showcasing the need to have hook-based deterministic access control approaches that are not based on prompt ingestion.

In \cite{mccauley2026HIBenchmarkJasonHiddenLayer}, a benchmark to measure how effective are models at enforcing conflicting instructions specified at different hierarchies (system vs. user), demonstrating that current models do not respect these hierarchies, and using \textit{Pretty Please} enforcement (prompts) is not enough to reliably enforce policies.

Agent Meltdowns under non-adversarial settings are studied in \cite{jha2026agent-metldowns-vitaly}, showcasing the need to monitor and include defenses outside the LLM.
The paper studies the behavior of agents when they encounter simple benign errors such as missing an expected file, and reports 64.7\% of agent rollouts that encounter simulated errors, resulted in dangerous actions such as conducting unauthorized reconnaissance or subverting access control. Interestingly, the study finds that in over half of the dangerous failures (meltdowns), the unsafe behavior was never reported to the user. These findings showcase the evident need of having defense in depth. 

The Coalition for Secure AI also published a set of recommendations to address agent containment challenges \cite{cosai-agent-insider-threat} (agents escaping their sandboxes), where using hooks to monitor and stop unexpected and malicious behavior was recommended urging treating agents as an insider threat. Overall, it requires having a more robust policy enforcement.

\subsection{Guardrails and Enforcement Frameworks} 
Guardrails models have been developed to prevent a variety of harms \cite{padhi2024graniteguardian,llamaguard,padhi2024graniteguardian,shieldgemma2,hap-small} by intercepting input-output responses of LLMs. Most of them inspect for pre-defined risks. Recently, some of them have added ways to verify for particular risks using techniques such as bring your own risk that can be used for policy verification\footnote{To see an example of such usage: \url{https://github.com/ibm-granite/granite.trust.policy-tools/blob/main/notebooks/guardian_enforcement.ipynb} }.  
Other approaches focus on multi-turn interactions.
A recent work monitors multi-turn conversation for 
\cite{guida2026cognitivefirewall} that monitors independently intent, context plausibility and manipulation and consistency. %
NeMo Guardrails \cite{rebedea2023nemo} is a toolkit to define and steer conversations based on pre-defined policies.
Valid conversation flows are specified in the formal language \textit{Colang}.
At runtime NeMo acts as a proxy between the user and the LLM,
selecting guardrails and modifying conversation flows according to Colang rules.
Enforcement of policies relies on selecting the next steps to guide the conversation.
When a user-model interaction is taking place, an \textit{embedding vector} is created and used to select the most relevant policy.
As such, policy conflict resolution is based on a stochastic nearest neighbor algorithm.
The final enforcement is also stochastic and uses prompting with in-context learning.  
This external verification adds defense in depth.

\subsection{Harnesses and Systems}
Multiple frameworks have been proposed in this space.
Mellea \cite{mellea} aims to \textit{``build predictable AI without guesswork"} by creating \textit{generative programs} written in python which verify compliance with requirements during the workflow of a GenAI application. Mellea uses \textit{hooks} \cite{mellea-hooks} to enforce policies linked to tool calling. A compiler to certify skills with Mellea was also provided \cite{mellea:compiler:2026}.
LangChain \cite{langchain} is a platform to build AI agents, where users can implement custom or predefined guardrails. Rails regulate pre-and-post tool calling interactions. 
Vertex AI Engine \cite{vertexAIEngineGoogle}, a product offering from Google enable the deployment of agents. The framework allows users to specify python functions, authenticate, access control and enable content filters. There is no default way to specify and manage conflicts caused by multiple policies.
CodeGuard from CISCO \cite{cisco-projectGuard} provides a list of rules to help prevent the generation of vulnerable software. 
These rules can be used before, during and after code generation to prevent insecure defaults, hard-coded secrets, use of outdated cryptographic algorithms. 
Open Claw \cite{open-claw} is an open-source autonomous artificial intelligence agent 
Agents specification is defined in a file SOUL.md that contains its name, desired tone, hard limits (never share private information) and core behaviors \cite{open-clau-soul}. However, these are essentially \textit{Pretty Please} prompts.
Enforcement mechanisms in these two clusters are usually found as packages, plugins or modules for users to integrate into their application \cite{cisco-projectGuard}.

\subsection{Policy Composition Conflict Resolution}
Policy conflict resolution has been studied extensively for traditional access control systems (see Section \ref{sec:policy-enforcement-principles}). 
However, enforcing policies in GenAI applications requires addressing conflicts of a variety of policies wider than access control (see Table \ref{tb:policy-enforcement-places}).
Work in this area mainly focuses on multi-agent policy agreement, where conflicts in policies among multiple agents are resolved using pre-defined rules, voting to arrive into a consensus or using an LLM to resolve the discrepancy among policies \cite{arionresearch2025}. 
Other approaches use \textit{Rule-based precedence systems} like OPA/Rego use explicit deny-overrides or permit-overrides semantics borrowed from traditional access control.
\textit{Semantic similarity approaches} e.g., \cite{rebedea2023nemo} select the most relevant policy based on embedding distance to the user request, but this introduces stochasticity into what should be a deterministic decision. 
None of these approaches provide formal guarantees when policies span multiple enforcement types (e.g., a soft content policy conflicting with a hard access control policy).

\section{Conclusion} \label{sec:conclusions}
    
The dramatic growth in capabilities of LLMs has led to new applications and the need to ensure compliance with policies. Our survey of what practitioners refer to as \textit{policy} reflects the fact that there are a big set of diverse policies that vary in their objectives and where they need to be enforced. 
We proposed dissecting policy techniques to identify good and bad practices by taking into consideration three different dimensions of policies and how they relate to each other.
Fig.~\ref{fig:takeaway-quadrants} 
summarizes our recommendations in terms of policy enforcement selection. 
In addition, we draw the following lessons:

\begin{enumerate}
    \item \textit{Just because you can, doesn't mean you should}
    Our methodological framework highlights that the capabilities of LLMs have led to a plethora of \textit{fuzzy} policy definitions that could not be enforced prior to the LLM era. At the same time, we see ``unreliable" enforcement mechanisms (LLM-based) being used to enforce strict policies that can be enforced with hard mechanisms. These deployment choices can lead to a false sense of security: if there is a well-defined policy, enforcement mechanisms that are non-stochastic should be used.

    \item \textit{Defense in Depth: A Sprinkle Donut is not the same as a Swiss Cheese model}
    Our analysis also shows an increase of prompt-based solutions that tend to encompass multiple desirable conditions to be enforced centrally by a single model.
    We argue that this \textit{Pretty Please Policy} enforcement mechanisms lead to a Sprinkle Donut not a Swiss Cheese Model: the same model cannot guard itself. True defense in depth requires having multiple mechanisms besides the LLM.
    This is particularly true given that LLMs are well known for their vulnerability to stochastic failures, jailbreaks, prompt injection attacks and reward hacking.  

    \item \textit{Stack Enforcement}
    Compliance requires logging information to verify the desired properties of a system.
    We argue that lessons learned by traditional cybersecurity are \textit{mostly} applicable to GenAI applications. This include the use of Policy Enforcement, Decision and Information Points while designing GenAI applications. Other solutions such as those inspired by SELinux hooks can help shape the field.
    In Section \ref{sec:policy-enforcement-principles} we discuss some nuances that make it difficult to fully apply some of the concepts to GenAI applications. These remain open challenges for the community to address.

    \item \textit{Interoperability and Conflict Resolution}
    Our analysis surfaces two gaps the community must address. First, standardized failure semantics across GenAI applications: the equivalent of HTTP 404 for policy violations. Some initial solutions~\cite{baracaldo:granite:policy} propose a unified set of exception codes well understood across GenAI applications.
    Second, principled approaches to resolving conflicts when multiple policies apply simultaneously. Current approaches rely on vector similarity, which lacks the formal guarantees of traditional access control conflict resolution. Addressing conflicting policies across the diverse set of policies a single application is subject to remains an open challenge.

\end{enumerate}

We hope that this paper leads to a constructive collective understanding of what different communities mean by policy and of the multiple efforts that aim to make GenAI safer, secure and compliant.
We proposed an analytical framework that offers a way for practitioners to systematically identify potential failures and gaps, and truly develop a defense-in-depth solution.

\newpage

\begin{small}
\section*{Acknowledgments}
I wrote this paper while preparing for my \textit{USENIX Security 2026 Enigma} track presentation \cite{baracaldo2026Enigma} as extended companion material of my talk. I want to thank Kendra Albert and David Freeman who listened to an earlier version of my presentation and provided feedback. Their input helped shape the presentation and analogy of the LLM as a \textit{donut} with a hole; originally it was another type of pastry. My points of view have been shaped by extensive reading, examining source code and talking to multiple practitioners. I would like to thank them all.
\end{small}

\begin{small} 
\section*{Author Biography}
\noindent \textbf{Nathalie Baracaldo}
 is a Senior Research Scientist and Master Inventor at IBM Research in San Jose, California, where her work focuses on building trustworthy AI systems. She has extensive experience delivering impactful machine learning solutions that are highly accurate, withstand adversarial attacks, and protect data privacy. Her current research focuses on safeguarding generative AI through unlearning and alignment techniques. She served as the principal investigator for the DARPA GARD program, leading efforts to extend and maintain the Adversarial Robustness Toolbox (ART) for red teaming evaluations. She also led IBM's federated learning initiative and co-edited two books: ``Federated Learning: A Comprehensive Overview of Methods and Applications" (Springer, 2022) and ``Machine Unlearning for Governance of Foundation Models" (2026). Her research has been published in top AI and security conferences, earning multiple best paper awards and thousands of citations. She received the IBM Master Inventor distinction in 2020 and the Corporate Technical Recognition in 2021. She holds a Ph.D. from the University of Pittsburgh.
\end{small}

\bibliographystyle{plain}
\bibliography{references}

\end{document}